\documentclass[preprint]{aastex}

\usepackage{natbib}

\newcommand{\ja}{JaFu~1}
\newcommand{\iras}{IRAS~18333}
\newcommand{\msun}{M$_\odot$}
\newcommand{\lsun}{L$_\odot$}

\newcommand{\kms}{km~s$^{-1}$}

\shorttitle{PNe central stars in Galactic globular clusters}
\shortauthors{Jacoby et al.}

\begin{document}


\title{Masses of the Planetary Nebula Central Stars in the Galactic Globular Cluster System from {\em HST} Imaging and Spectroscopy\footnote{Based, in part, on observations made with the NASA/ESA Hubble Space Telescope, obtained [from the Data Archive] at the Space Telescope Science Institute, which is operated by the Association of Universities for Research in Astronomy, Inc., under NASA contract NAS 5-26555. These observations are associated with program GO-11558.}}


%% Note that the \email command in aastex V6 positions the email info strangely.
%% Greg suggested altaffil but I chose to append to author address 

\author{George H. Jacoby}
\affil{\textnormal{Lowell Observatory, Flagstaff, AZ 86001; gjacoby@lowell.edu}}
%\email{gjacoby@lowell.edu}

\author{Orsola De Marco\altaffilmark{1,2}}
\affil{Department of Physics and Astronomy, Macquarie University, Sydney, NSW 2109, Australia; orsola.demarco@mq.edu.au}
%\email{orsola.demarco@mq.edu.au}

\author{James Davies}
\affil{Space Telescope Science Institute, Baltimore
MD 21218; jdavies@stsci.edu}
%\email{jdavies@stsci.edu}

\author{I. Lotarevich}
\affil{American Museum of Natural History, New York, NY}
%\email{ilotarevich@yahoo.com}
%\email{llah6am@gmail.com}

\author{Howard E. Bond}
\affil{Department of Astronomy \& Astrophysics, Pennsylvania State University, University Park, PA 16802; heb11@psu.edu}
%\email{bond@stsci.edu}

\author{J. Patrick Harrington}
\affil{University of Maryland, College Park, MD; jph@astro.umd.edu}
%\email{jph@astro.umd.edu}

\and

\author{Thierry Lanz}
\affil{Laboratoire Lagrange, Universit\'e C\^ote d'Azur, Observatoire de la C\^ote d'Azur, CNRS, 06304 Nice, France; thierry.lanz@oca.eu}
%\email{thierry.lanz@oca.eu}


\altaffiltext{1}{Astronomy, Astrophysics, and Astrophotonics Research Center, Macquarie University, Sydney, Australia}

\altaffiltext{2}{Formerly at the American Museum of Natural History, New York, NY. Research associate at the American Museum of Natural History, New York, NY.}


\begin{abstract}
The globular cluster (GC) system of our Galaxy contains four planetary nebulae (PNe): K~648 (or Ps~1) in M15, IRAS 18333-2357 in M22, JaFu~1 in Pal~6, and JaFu~2 in NGC~6441. Because single-star evolution at the low stellar mass of present-epoch GCs was considered incapable of producing visible PNe, their origin presented a puzzle.
We imaged the PN JaFu~1 with the {\em Hubble Space Telescope (HST)} to obtain photometry of its central star (CS) and high resolution morphological information. We imaged IRAS~18333-2357 with better depth and resolution, and we analyzed its archival {\em HST} spectra to constrain its CS temperature and luminosity. All PNe in Galactic GCs now have quality {\em HST} data, allowing us to improve CS mass estimates.
%% deleted for v8: through the core-mass vs luminosity relation. 
We find reasonably consistent masses between 0.53 and 0.58 \msun\  for all four objects, though estimates vary when adopting different stellar evolutionary calculations. The CS mass of IRAS~18333-2357, though, depends strongly on its temperature, which remains elusive due to reddening uncertainties.
For all four objects, we consider their CS and nebula masses, their morphologies, and other incongruities to assess the likelihood that these objects formed from binary stars. Although generally limited by uncertainties ($\sim0.02$ \msun) in post-AGB tracks and core mass vs luminosity relations, the high mass CS in K~648 indicates a binary origin. The CS of JaFu~1 exhibits compact bright [O~III] and H$\alpha$ emission, like EGB~6, suggesting a binary companion or disk. Evidence is weaker for a binary origin of JaFu~2. 

% We also question the classification of IRAS~18333-2357 as a PN due to numerous anomalies in its properties. 

% GEORGE - Deleted the final sentence in the abstract because it exceeds the word count limit, and is alluded to earlier.
%\textnormal{Using a recent, completely new set of stellar evolutionary tracks, the stellar masses could be lower but the evolutionary timescales would be significantly at odds with the PN kinematic ages, a discrepancy that could be reconciled with binary origins.}

\end{abstract}





\section{Introduction}

According to conventional single star stellar evolution models \citep{Schoenberner1983}, the old stellar populations of Galactic globular clusters (GCs) are predicted to be unable to host planetary nebulae (PNe). In such old populations, the masses of stars that are today transiting between the asymptotic giant branch (AGB) and the white dwarf (WD) phase should be smaller than $\sim$0.55~\msun\  and measurements of WD masses in GCs support this expectation. \citet{Kalirai2009}, for example, find that GC stars evolving at the present epoch typically produce 0.53~\msun\ WDs. At these low masses, post-AGB stars are unable to develop a visible PN because the transition time, the time it takes a post-AGB star to warm up to temperatures capable of ionizing oxygen and/or hydrogen, is too long. By the time the star is hot enough ($>10^5$ yrs), the mass lost during the preceding AGB evolution will be too dispersed ($\sim10^{4-5}$ yrs) to make a detectable PN. 

Recently, \citet{MillerBertolami2016} presented arguments for much faster transition times, between 25,000-75,000 yr for 9-12 Gyr GC stars, thereby enabling slightly \textnormal{less massive, older} GC stars to produce a PNe.
%younger GC stars to produce PNe. 
%: not younger.... I would say "slightly less massive, older GC stars to produce a PNe."
For the old clusters discussed here (11.6-13.2 Gyrs), those shorter times are either marginal or too long for a detectable PN to be formed, and \textnormal{are several} times larger than the kinematic ages of the PNe discussed here.

In a survey of $\sim$133 GCs, \citet{Jacoby1997} detected four PNe where two were previously known (Ps~1 / K~648 in M15 and IRAS~18333-2357 in M22). All four are confirmed members, based on the close correspondence to their host cluster radial velocities, extinctions, chemical compositions, and spatial coincidences. Similarly, \cite{Jacoby2013} and \cite{Bond2015} report \textbf {only} a small number of PNe found in M31 and throughout the Local Group, consistent with the rates in the Galaxy. 

Because the existence of these PNe violates the conventional stellar evolution predictions, and because of the statistical association of the host clusters with X-ray brightness, \citet{Jacoby1997} suggested that some form of binary interaction was responsible for their evolution. Shortly thereafter, \citet{Alves2000} proposed mechanisms for binary stars to produce PNe in GCs. And in their detailed discussion of the nature of Ps~1, \citet{Otsuka2015} also argue that it descended from a coalesced binary.

Currently \textnormal{three} hypotheses can explain the existence of PN in globular clusters. \textnormal{First, there is the recent revision to post-AGB timescales presented by \citet{MillerBertolami2016}, in which normal globular cluster stars are, in fact, able to form PNe. However, in this hypothesis where normal GC stars go through a PN phase, we expect to see up to five times more PNe in the Galactic cluster population \citep{Jacoby1997}}.

\textnormal{Second,} \citet{Ciardullo2005} hypothesized that \textnormal{those PN central stars with masses larger than the cluster WD mass} are the descendants of blue straggler stars (BSS). BSS have abnormally high main sequence masses for their cluster age and are the supposed descendants of stellar mergers or binary mass transfer \citep{McCrea1964}. In such a case the PN would be indistinguishable from one created by a single star of larger mass. A slight variation on this scenario is that the star suffered a merger during either of the giant phases. The merged star would then evolve as a single, more massive object and make a normal PN.

Alternatively, PNe \textnormal{may have central star masses commensurate with the cluster WD mass. But if their nebular kinematic ages are too short for the corresponding evolutionary timescales, they can still be explained as post-common envelope binaries.} The common envelope phase transforms a giant into a much more compact object in a matter of years \citep{Sandquist1998,Passy2012}. The post-common envelope primary, however, has the same luminosity as its giant progenitor because the luminosity is related only to the core mass, which has not changed. Thus, a common envelope effectively transforms a cool giant into a hot central star in a time commensurate with the giant's dynamical timescale, thus shortening
the time of transition. The exact temperature depends on the orbital separation after the common envelope, because this determines how large the star can be. As a result, a post-common envelope PN kinematic age should always be smaller than the age of the central star implied by its effective temperature and mass. Conversely, if trying to obtain the mass of the central star using the age of the nebula, we would systematically obtain a larger mass.

The difference between \textnormal{these} scenarios is that the masses of mergers would be naturally larger than the typical WD masses in the cluster, while if the PNe are produced in a common envelope, the masses of the central stars need not be larger than the WD masses at the top of the cooling sequence, and can even be smaller. \textnormal{In that case, there would be a large discrepancy between the kinematic age of the nebula and the post-AGB age of the central star}.  

\textnormal{In general for the GC PNe, though, the central star masses cannot be derived with sufficient accuracy to discern a binary merger origin from other scenarios solely from the mass. For this paper, we compare results obtained from different stellar evolutionary calculations and we also utilize the {\em HST} images to assist in discriminating a binary origin from other possibilities.}

Irrespective of any particular scenario that explains the origin of these four PNe, they are of interest in that they are a group of closely related but unusual PNe at known distances, possibly the result of mergers, a class of star that is otherwise difficult to identify. With the goal of clarifying their evolutionary status, we have obtained  {\em HST}  observations of two PNe:  JaFu~1 in Pal~6 and IRAS~18333-2357 (hereafter IRAS~18333) in M22 (NGC~6656). We have supplemented these data with pre-existing  {\em HST}  images and photometry of JaFu~2 in NGC~6441 and of Ps~1 in M15 (NGC~7078).

With these data, plus improved cluster properties from the literature (e.g., distances, ages, and
compositions), we are able to derive improved estimates for the 
PN central star luminosities, temperatures, and consequently, 
masses. In addition, the data set provides a clearer picture of 
the nebular morphologies. The more complete picture of these objects allows us to be more definitive about the likelihood that these central stars are the descendants of binary interactions.

\section{Observations and Data Reduction}

Of these four PNe, only \ja\ had not been observed previously with {\em HST}. The archival data quality for the other three PNe, though, is not all of comparable quality, and the image of IRAS~18333 is especially lacking in depth. Thus, we obtained new  {\em HST}  imaging data for both \ja\ and IR~18333 using either the Wide Field Planetary Camera~2 (WFPC2) or the Wide Field Channel of the Advanced Camera for Surveys (ACS/WFC) on  {\em HST}  under program GO-11558 (PI: De Marco). A journal of the observations is given in Table 1.  {\em HST}  FOS UV spectra of the central star of \iras\  obtained in 1994 were collected from the archive for the program by \citet{Harrington1996}.

\begin{figure}
\includegraphics[width=0.95\textwidth,natwidth=610,natheight=642]{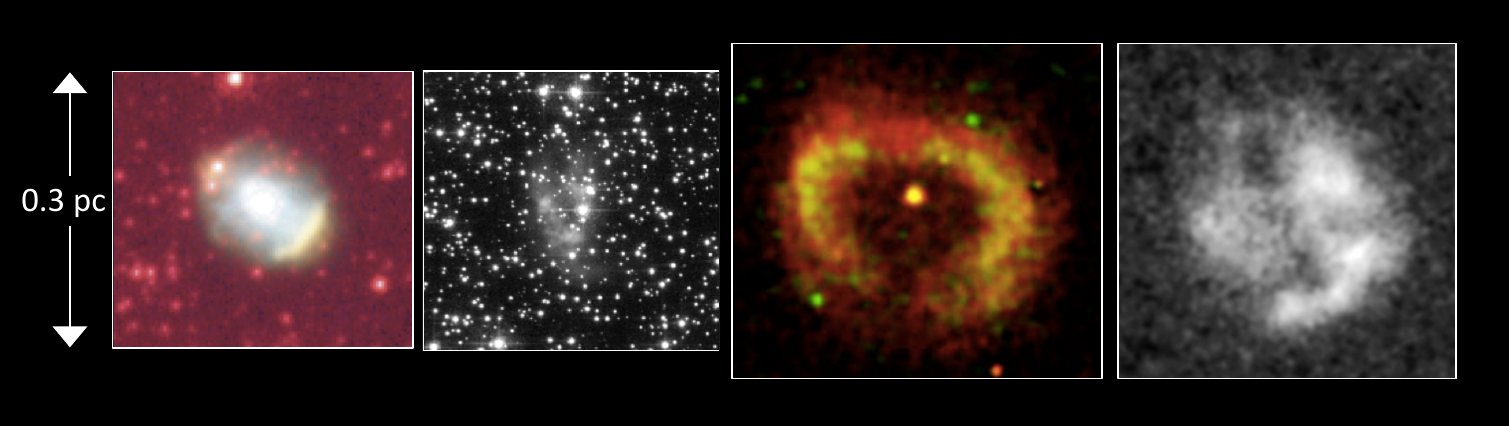}
\caption{The four Galactic globular cluster PNe placed on a common physical scale. From left to right: Ps~1 (K648) is taken from \citet{Alves2000} where the  orientation has north 11 deg clockwise from the right and east is 11 deg clockwise from up, and colors are [O~III] (blue), H$\alpha$ (green) and [N~II] (red). From this paper, IRAS~18333 is shown in [O~III], while JaFu~1 is shown in [O~III] (green) and H$\alpha$ (red). JaFu~2 in shown in H$\alpha$, taken from \citet{Jacoby1997}. North is up and east is to the left with the exception of Ps~1. Note the non-classical PN morphology of IRAS~18333, and the similarity between JaFu~1 and the Necklace Nebula \citep{Corradi2011}.\label{fig:GC_PN_Scaled}}
\end{figure}

Figure~\ref{fig:GC_PN_Scaled} illustrates the images of all four GC PNe at the same physical scale. Their sizes are all comparable, being between 0.11 and 0.26 pc in diameter. On the other hand, their morphologies are diverse as discussed later.

\section{The Expected White Dwarf Masses for Globular Clusters}

The ages and metallicities of the clusters  are given in Table 2. Based on these and cluster models, for example from \citet{Sippel2013}, we expect the turnoff masses to be $\sim0.80 - 0.84$ \msun\ using the \citet{Bertelli2008} stellar evolutionary tracks for a metallicity $Z=0.004$.
%, but we note that this estimate would not be very different for solar metallicity. 
%  0.85 for solar instead of 0.84

In principle, we can apply the initial-to-final mass relation (IFMR) to predict the expected CSPN mass in a cluster. We would then compare the derived masses of the central stars in the clusters with the predicted masses to determine if the actual central stars are consistent with single star evolution, or if they require some mass enhancement. 

Unfortunately, the IMFR is not precisely defined for initial masses as low as those in GCs. In this regime, predicted core mass differences of 0.01-0.02~\msun\ imply significant deviations in their luminosities and evolutionary timescales. For example, if we use the observationally-determined IFMR by \citet{Kalirai2009} where $M_{\rm final} = (0.101 \pm 0.006) M_{\rm initial} + (0.463 \pm 0.018)$ (see also Kalirai 2012), we derive a predicted range in final masses for our clusters of 0.54--0.55~\msun. Similarly, at its lowest end, \citet{Weidemann2000} observationally derived an IFMR that indicates a 1.0~\msun\ main sequence star produces a 0.55~\msun\ WD, and so an old cluster WD would be even less massive. We have also independently applied the solar metallicity IFMR calculated in \citet{DeMarco2011} from the MESA code by \citet{Paxton2011} to obtain a range of 0.50-0.51~\msun\ for a more metal-rich old cluster. 

% PARAGRAPGH DELETED AT SUGGESTION OF REFEREE:
% Alternatively, we can fit the \citet{Bertelli2008} tracks to derive a WD mass of 0.56~\msun\  for a Z=0.004 old cluster. \citet{Buell2012} also used the \citet{Bertelli2008} tracks along with a code derived from \citet{Renzini1981} to treat the TPAGB and he obtained a range of 0.52-0.53~\msun\ for Z=0.004 and Y=0.23. We also note that he derived 0.56-0.57~\msun\ if a higher helium abundance of Y=0.33 is adopted. 
% We cannot understand  that paper why the predicted CSPN masses are lower than the corresponding core masses at the beginning of the thermally-pulsating AGB as there is no comment in their section 2.8 on how the core is expected to decrease in mass during the TPAGB. \citet{Meng2008} claim a higher final mass at lower metallicity: using their fitting formulae, final masses of 0.58~\msun\ and 0.55~\msun\ are derived for Z=0.004 and Z=0.02 respectively. Finally, \citet{PradaMoroni2007} predict a WD mass $\sim$0.57~\msun\ for a turnoff of 0.80-0.84~\msun\ for Z=0.004.

Thus, the  range of predicted central star masses for GC central stars extends from 0.50 to \textnormal{0.55~\msun}, which is so wide that a direct comparison with observed CSPN masses is compromised. This range in core mass corresponds \textnormal{to heating} timescales between $\sim 10^6 - 10^4$ years. We therefore adopt the observationally measured WD mass of $\sim0.53\pm0.01$~\msun\ as the expected value of a GC central star mass from \citet{Kalirai2009} for the GCs M4, NGC~6397, 47~Tuc, and $\omega$ Cen. These measurements fall in the mid-range of the above predictions, and are further supported by the \citet{Woodley2012} observations of 47~Tuc, and by the \citet{Hanson2004} observations of M4. This value is also higher than the average WD mass for the cluster M4 and NGC~6572 determined by \citet{Alves2000} ($\sim0.50\pm0.02$~\msun) based on the M4 WDs (0.52$\pm$0.03~\msun), and on the re-calibrated WD masses for the cluster NGC~6572 (0.46$\pm$0.04~\msun). 
We presume that the more recent work of \citet{Kalirai2009} supersedes the considerations of \citet{Alves2000}. 

\textnormal{Furthermore, not only are the expected stellar remnant masses uncertain, but the resulting post-AGB luminosities of GC stars are subject to several factors that may not be known or modeled accurately. These include the opacity, molecular weight, nuclear reaction rates, chemical compositions, and stellar temperature \citep{Marigo2000}. Thus, there is no simple core mass-luminosity relation to adopt for deriving a central star mass from a given luminosity. Also, post-AGB models for these low mass, low metallicity stars are generally unavailable and extrapolation of various existing models may be necessary. Consequently, the great advantage offered by GCs, where we can know the distance and luminosity of a CSPN quite well, is offset by the uncertainties in converting the luminosity into an accurate mass.}

\section{The Central Star Masses}

In this section, we review the available information on the four central stars and revisit the mass determinations found in the literature.

\subsection{K~648, the central star of Ps~1}

Several groups have derived masses for K~648, the central star of Ps~1. \citet{Alves2000}, for example, determined a mass of $0.60\pm0.02$~\msun\ by referring their derived CS luminosity (6000~\lsun) and temperature (40,000 K) to the core mass - luminosity relation of \citet{Vassiliadis1994} where $L=56694$ ($M_c$/\msun\ - 0.5). This relationship is an approximation that clearly fails for WD masses near or below 0.50~\msun\, but it is sufficient for this object. The observed brightness and temperature were based on extensive multi-wavelength photometry and spectral absorption-line fits, and an adopted distance of 12.3 kpc. When we adjust the luminosity for the smaller, more accurate, distance of $10.0\pm0.5$ kpc \citep{McNamara2004} \textnormal{to $\sim4000$~\lsun}, the CSPN mass decreases to $\sim$0.575~\msun. 

\citet{Rauch2002} derived a spectroscopic mass of 0.57$^{\rm +0.02}_{\rm -0.01}$~\msun\ using a $\log g=3.9$ and $T_{\rm eff}$=39,000 K which independently suggests a distance of 11.1~kpc. And, \citet{Bianchi2001} obtained a spectroscopic mass of 0.62$\pm$0.10~\msun\ from a $\log g$=4.0, $T_{\rm eff}$=37,000 K, and a luminosity of $\log L$=3.45, assuming a distance of 10.3~kpc. 

In a recent and thorough analysis, \citet{Otsuka2015} re-evaluated the available  {\em HST}  photometric data, added high resolution spectroscopy from Subaru, and a diverse array of multiwavelength data and modeling to derive the PN properties of Ps~1. They derive $\log g$=3.96, $T_{\rm eff}$=36,360 K, similar to previous values, leading to a central star mass of at least 0.61~\msun\  for a distance of 10.3~kpc. This corresponds to an initial main sequence mass of at least 1.15~\msun, which is far too large to derive from a GC main sequence star, leading \citet{Otsuka2015} to argue that this object formed from a binary star merger.  

\textnormal{\citet{MillerBertolami2016} recently revised the stellar evolutionary tracks for post-AGB stars with updated microphysics. The hydrostatic equilibrium structure of a star changes sufficiently that departure from the AGB takes place at a different envelope mass and results in overall lower core masses for the same luminosity, as well as faster post-AGB evolutionary timescales for a given core mass.}

\textnormal{Thus, for the K~648 luminosity of 4000~\lsun\  and interpolating between the Z=0.001 and Z=0.0001 evolutionary tracks, we obtain a mass of 0.53~\msun. A post-AGB star with this mass takes approximately 24,000~yrs\footnote{The timescales are calculated from the end of the AGB, defined by Miller-Bertolami (2016) as the time when the envelope mass decreases to 1 per cent of the stellar mass.} to reach 37,000~K, the temperature at which a star will ionise O$^+$ to produce [O~III] and is about the temperature of K~648. While a star of this low core mass could ionise the nebula before it dissipates, that timescale is inconsistent with the carefully calculated nebula kinematic age of $\sim5700$ yrs \citep{Schoenberner2014}. One way to explain this discrepancy is if the AGB star suffered a common envelope phase, forcing rapid ejection of the envelope and heating of the central star. In this scenario, the central star would be a close binary.} 

\textbf{Alternatively, if we were to adopt the \citet{Weiss2009} tracks, then the central star could have a mass of 0.53~\msun but with a much faster evolutionary timescale ($\sim10,000$ yrs) and thus, would be consistent with a single star origin.}

% DELETED as redundant with previous paragraph: 
% Unless the distance to M15 is much smaller than the evidence suggests, the core mass of K648 is well above the adopted GC white dwarf mass of 0.53~\msun\ (Section 3).

%----------------
\subsection{The central star of IRAS~18333}

Deriving the mass of this star is complicated by the difficulty in measuring its temperature. Several authors have constrained the temperature range, but the uncertainty remains large. This is due, in part, to the uncertain reddening estimate, and also due to the poor quality of the available spectra.

\citet{Gillett1989} noted that the optical spectrum of the CS is similar to that seen in planetary nebulae, suggesting that the temperature must be greater than 35,000~K and could be as high as 80,000~K. Recognizing that the temperature is critical to understanding the nature of this object, \citet{Cohen1989} used IUE spectra and ground-based echelle data to constrain the temperature. The presence of He~II absorption and the absence of He~I absorption argued that the star must be hotter than 50,000~K.

Subsequently, \citet{Pena1992} combined the temperature of 50,000~K from \citet{Cohen1989} with the visual magnitude and reddening from \citet{Gillett1989} to derive a mass of 0.56~\msun.

One can also fit the continuum flux by a black-body, but this approach suffers some degeneracy due to the uncertain reddening. Adopting $E(B-V)=0.50$ from \citet{Cudworth1990} yields a temperature of $\sim$70,000~K, which leads to a very high mass as discussed below.

\citet{Harrington1993} obtained high signal-to-noise echelle optical spectra noting again that the temperature must be greater than 50,000~K because there is no He~I $\lambda$4471 absorption. Taking a somewhat tangential approach to the problem, they identified a 75,000~K star having similar characteristics and for which a detailed NLTE model had been developed by \citet{Rauch1991}. At this temperature, \iras\  has a luminosity of 13,650~\lsun, which corresponds to a mass of ~0.75~\msun.  This high mass has been commonly adopted in the literature, thereby placing \iras\ in an unusual position among the general PN population, and is unique among PN central stars in GCs. A central star as massive as 0.75~\msun\  implies a progenitor mass exceeding 3~\msun, thereby requiring the unlikely merger of at least four globular cluster stars, and consequently arguing indirectly for a lower temperature.

\begin{figure}
%\vspace{10cm}
\includegraphics[viewport= 0 100 640 700,width=1.2\textwidth,natwidth=685,natheight=1100]{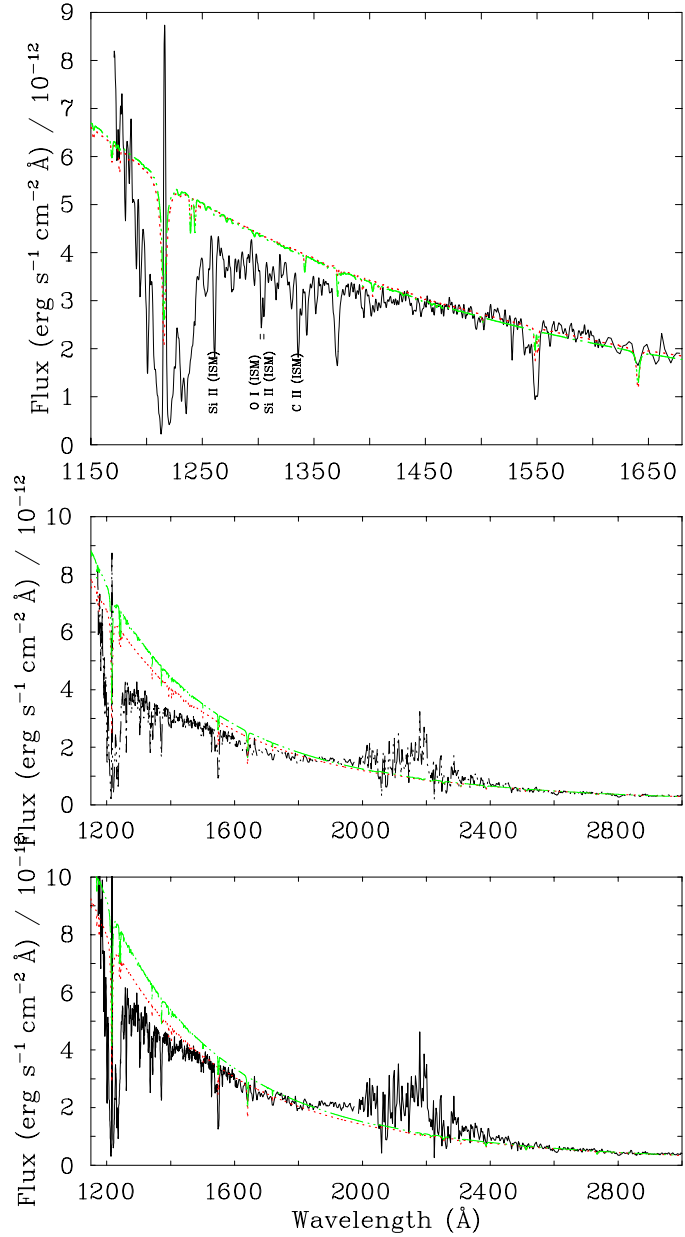}
%\special{psfile="IRAS18333Fitting-Hnormal.pdf"}
\caption{A fit to the observed (solid black line) UV spectrum of IRAS18333 using hydrogen-normal models with 50,000~K (red line) and 70,000~K (green line) synthetic atmospheres. Top: the models are normalised at $\sim$1450~\AA; Middle panel: the models have been normalised at $\sim$3000~\AA; Bottom panel: the models have been normalised at 3000~\AA, but the data have been dereddened assuming $E(B-V)$ = 0.54~mag.\label{fig:IRAS18333-spectrum}} 
\end{figure}

We have re-evaluated the arguments for this high mass, starting with a re-examination of the  {\em HST}  UV spectrum obtained under program GO-5690 (PI: Harrington). Figure~\ref{fig:IRAS18333-spectrum} illustrates this spectrum, along with the IUE spectrum obtained by \citet{Cohen1989}. We have dereddened the spectra by $E(B-V)=0.5$~mag and spliced the FOS and IUE spectra. We also shifted the spectrum to the velocity of the cluster ($-146$ \kms; Harris 1996 [2010 edition]). The strong absorption lines of Si~II at 1360 and 1304 \AA, of O~I at 1302~\AA, and of C~II at 1335~\AA\ are interstellar in nature.

We fit numerous NLTE line blanketed stellar atmosphere models using the TLUSTY code \citep{Hubeny1995, Lanz2003}, assuming temperatures between 50,000~K and 70,000~K, $\log g $= 5.5 or 6.5 and a metallicity of 3\% solar,  but none of the models fits the continuum well. By increasing the reddening to raise the continuum, especially in the UV, we obtain a better fit with $E(B-V)$ = 0.54. But even with the lower temperature of 50,000~K the model fit of the continuum is poor with the observed flux being too low between 1300 and 1400~\AA. Additionally, the larger reddening is inconsistent with the strength of the 2200~\AA\ feature, though, we cannot discount the presence of nebular dust having different properties from those assumed for interstellar dust \citep{Cardelli1989}.

Unfortunately, the spectral lines do not help discriminate clearly between the two temperatures. The C~IV line at 1550~\AA\ is too weak in both models. The He~II line at 1640~\AA\ fits better than the C~IV line in either model. The O~V line at 1371~\AA\ is poorly fit and increasing the temperature does not provide a better fit. This line is likely formed by \textnormal{X-rays} in the wind of the star \citep{Herald2011, Guerrero2013}.

%http://adsabs.harvard.edu/abs/2011MNRAS.417.2440H

\begin{figure}
%\vspace{10cm}
%\special{psfile="IRAS18333Fitting-HdefLines.pdf"}
\includegraphics[viewport= 0 250 800 600,width=1.4\textwidth,natwidth=610,natheight=342]{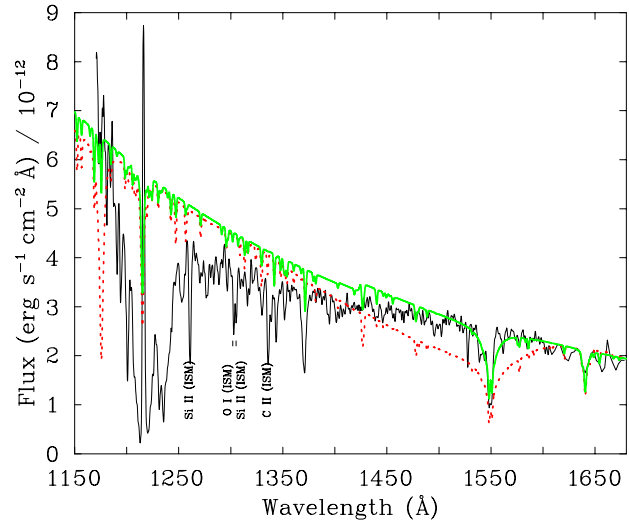}
\caption{A fit to the observed (solid black line) UV spectrum of IRAS~18333, dereddened by $E(B-V)$ = 0.5~mag, using hydrogen-deficient models with 50,000~K (red line) and 70,000~K (green line) synthetic atmospheres. \label{fig:IRAS18333-spectrum2}}
\end{figure}

The poor line fits and the hydrogen-deficient nature of the PN suggest that the star may be hydrogen-deficient and have an abundance pattern close to that of PG1159 stars \citep{Werner2006}. We therefore tried a set of abundances in line with those of PG1159 stars (He:C:O = 0.45:0.41:0.14, by mass). In Figure~\ref{fig:IRAS18333-spectrum2} we show the same fits as in Figure~\ref{fig:IRAS18333-spectrum}, but this time using the hydrogen-deficient models.

While the continuum issue remains unresolved, the line strengths are better matched. The C~IV line strength is now well fit by the hotter model and the He~II line remains reasonably well fit while the C~III line at 1175~\AA\ is far too strong in the cooler model. 
% Although fitting this line right at the edge of the spectrum and on the wing of the strong geocoronal absorption feature is risky, we can say with certainty that the observed line is not as strong as in the 50,000~K model. 
The strength of the line in the 70,000~K model is more commensurate with the observed one. The O~V line at 1371~\AA\ remains poorly fit even with the increased oxygen abundance, although the hotter model fits it better than the cooler one.

%It is difficult to select the best model, either based on the continuum fit or the line fits. The comparison of the fits to the C~III and C~IV lines does not provide an indication of temperature, mainly because most of the carbon is in the C$^{4+}$ ionisation state. At a $\log g$=6.5 the cooler, 50,000-K model, does have a significant abundance of carbon in lower ionization states. However, the already too strong C~III $\lambda$1175 line may be a result of a large carbon abundance, or an observational issue at the edge of the spectrum rather than an indication that the model is too cool. 

%Finally we compare by eye our spectral models to the optical spectra presented by \citet{Gillett1989}. Their spectrum of the South star, thought to be the central star on the grounds of its blue colour, shows some Balmer lines. They point out, however, that there may be some contamination from the north star and that they could not tell whether the Balmer lines are in fact He~II lines because of the poor spectral resolution.

We also compared the 50,000~K and 70,000~K models with normal and PG1159 compositions to the optical spectrum in Figure 4b of \citet{Gillett1989}. The carbon-rich models were compared to an approximate measurement of the optical spectra shown in the figure of \citet{Gillett1989}. We observe the following. (i) The data have deeper lines than the model on average, arguing against contamination of the continuum by another source such as the nearby ``North'' star. (ii) The 50,000~K model has lines whose average strengths are closer to the data than the 70,000~K model.  (iii) Both models, but the 50,000~K model more so, have more lines than just the He~II lines, while the data have essentially only the He~II lines (we assume that the Balmer lines are from He~II). (iv) The relative strength of the lines in both models are approximately the same as in the data. While not a temperature indicator, this fact further suggests that the Balmer lines are indeed He~II lines; hydrogen-normal models of the same temperature have hydrogen lines that are much stronger than the He~II lines.

In addition, we calculated a 55,000~K helium-rich model (no carbon) and the total number of optical absorption lines is closer to the data, but the He~II lines are too shallow. This suggests that, as suspected from the UV fits, a PG1159 carbon abundance may be too high. This exercise further supports the assertions that the star is unlikely to be H-rich and that the carbon abundance is likely lower than the value that we assumed in the PG~1159 models. Critically, the stellar temperature remains somewhat unconstrained between ~50,000~K and 70,000~K.

Lacking clear evidence that the star is as hot as 70,000~K or as cool as 50,000~K, we accept a mass uncertainty. The extinction-corrected luminosity in the case of the 50,000~K star is then $\sim3300$~\lsun\ and the central star mass is $\sim0.56$~\msun\ using the mass-luminosity relation of \citep{Vassiliadis1994} and the temperture-bolometric correction relation (eqn. 6) of \citet{Vacca1996}. Not surprisingly, this mass is very close to that found by \citet{Cohen1989} and \citet{Pena1992} and only slightly higher than that expected for a WD in a cluster. It is also nearly identical to what we find for the masses of the CSPN in JaFu~1 and JaFu~2. On the other hand, if the stellar temperature is as high as 70,000~K, then the luminosity would be $\sim8400$~\lsun\ and the mass would be 0.65~\msun, far too high to evolve from a single GC star. \textnormal{We add the further caution that a hydrogen deficient CSPN, that is therefore helium burning, requires an even greater mass to achieve an observed luminosity, according to \citet{Vassiliadis1994}.}

\textnormal{As we did for Ps~1, we also consider using the evolutionary tracks of \citet{MillerBertolami2016} for Z=0.001, appropriate for M22. We find central star masses of 0.54 or 0.59~\msun\  using the low and high effective temperatures respectively. The evolutionary timescales needed to reach the two temperatures are 35,000 and 1100 yrs, respectively. For this object, though, the simple apparent kinematic age of the nebula (i.e., radius divided by expansion velocity from \citet{Pereyra2016}) is $\sim4000$ yrs, which is intermediate between the two stellar evolutionary timescales. Thus, if the effective temperature is closer to 70,000~K, the star would have a higher mass and we would have to conclude that it suffered a merger. Its kinematic age in that case may be reconciled with the timescales expected for post-AGB evolution. On the other hand, if the temperature and mass are at the lower end of the possible range, the highly discrepant kinematic and post-AGB evolutionary ages would suggest that a common envelope history may have occurred.}

Thus far, we have assumed that the central star is a single degenerate star. A second degenerate star may be present as for PN~135.9+55.9 \citep{Tovmassian2010}, thereby confusing the temperature analysis, and consequently the mass estimate. If this is the case, we expect to see evidence of a binary, either through multiple line sets displaced by a velocity offset, or by velocity variations. Those expectations, though, could be compromised if the viewing angle is nearly pole-on or the stellar separation is sufficiently wide that existing spectra \citep{Cohen1989}, having a velocity resolution of $\sim15$ \kms, have insufficient resolution. This explanation is purely speculative at this point, but high resolution spectra would provide a definitive test.

%--------------------
\subsection{The central star of JaFu~1}

\citet{Jacoby1997} estimated the central star mass of JaFu~1 from the predicted temperature (82,000~K) and luminosity (570 L$_\sun$) implied from a Cloudy photoionization model \citep{Ferland1998}. These values lead to a mass of $\sim0.54\pm0.01$ \msun\ (for a distance of 5.9~kpc). The central star, though, had not been observed at that time, and so there was no check on the visual magnitude predicted by the photoionization model. The  {\em HST}  observations reported here provide that check.

We have reviewed the distance estimates to Pal~6, which were rather uncertain when \citet{Jacoby1997} performed their calculations. More recent and better founded measurements argue for a distance of 7.2~kpc \citep{Lee2002}, which we adopt here. 

The reddening was measured by \citet{Jacoby1997} from the nebular Balmer line decrement, H$\alpha$/H$\beta$ to be $E(B-V) = 1.93\pm0.12$. However, the reddening may be dependent on the exact line of sight, and there may be clumpy dust in the nebula. Rather than simply adopting this reddening value, we derive the extinction from the photometric colors of the central star. This approach has different systematics than the Balmer decrement approach, and therefore serves as a useful check.

At these large reddening values, the conversion from the  {\em HST}  WFPC2 magnitudes to the Vega system incurs a significant shift in filter effective wavelengths that depend on the reddening. The process to determine the reddening is therefore iterative but converges quickly.

There is an additional complication: because the central star hosts an unresolved emission-line source (see below), we must subtract the component of the F555W flux that is contributed by that source. This fraction is $\sim$10\%.
After correction, the implied $V - I$ color is approximately $-0.68$.

The intrinsic $V-I$ for a single star having a temperature of 82,000~K is approximately $-0.33$ \citep{DeMarco2013}. This value is 0.35 mag redder than the  {\em HST}  photometry, suggesting that the reddening from \citet{Jacoby1997} is too high. That conclusion is also indicated by the average cluster reddening of $E(B-V)=1.30$ derived by \citet{Lee2002}. \textnormal{Consequently,} either the nebula material contains significant amounts of local dust, the adopted reddening law is overcorrecting the observed magnitudes, or the interstellar reddening is patchy across the cluster. For the purposes of this paper, we assume the latter solution; a value of $E(B-V) = 1.67$ resolves the discrepancy and is intermediate between the values from \citet{Jacoby1997} and \citet{Lee2002}.

Given the  {\em HST}  V-band magnitude of 23.12, the distance, and the reddening, we find that M$_V$=3.65. This value is within 0.1 mag \textnormal{of the value} derived by \citet{Jacoby1997} using a photoionization model, where the temperature was predicted at 82,000~K. A bolometric correction of $-5.94$ is obtained by extrapolating the relation $BC = 27.66 - 6.84 \log T_{\rm eff}$ from \citet{Vacca1996}, leading to a
bolometric magnitude of $-2.29\pm0.15$. This corresponds to a bolometric luminosity of  $648\pm96$~\lsun. The implied central star mass is $0.53\pm0.02$~\msun\  based on \textnormal{an uncertain extrapolation to} the evolutionary tracks from \citet{Vassiliadis1994}. The mass uncertainty includes the potential error introduced by extrapolating the BC correction relation beyond 50,000~K to 82,000~K. 

\textnormal{We note that such low CSPN luminosities, combined with a moderately high temperature, are not consistent with standard post-AGB stellar evolution. The \citet{MillerBertolami2016} tracks for Z=0.001 metallicity indicate that the star would have a core mass of 0.53~\msun\ and it would be on the cooling track with a temperature of 117,000~K. This is much higher than the photoionization model temperature of 82,000~K. Furthermore, the timescale to reach that point would be 78,000 yrs, which is inconsistent with the simple kinematic age estimate of 8,000 yrs (using the expansion velocity from \citet{Pereyra2016}).} 

\begin{figure}
\includegraphics[width=1.0\textwidth,natwidth=610,natheight=642]{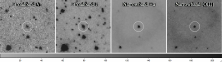}
\caption{A comparison of the images of JaFu~1 in four different bands. By comparing the brightness of the central star of JaFu~1 to nearby stars, we see that both H$\alpha$ and [OIII] images are relatively overluminous.\label{fig:BandComp}}
\end{figure}

Our  {\em HST}  observations revealed an \textnormal{additional} anomaly. The central star flux brightens significantly in the emission-line bandpasses. Figure~\ref{fig:BandComp} illustrates the relative brightness of the central star of JaFu~1 in the four bands. The luminosity of the central star in [O~III] and H$\alpha$ is $\sim10$ times higher than expected from the V and I fluxes, suggesting the presence of an unresolved emission-line nebulosity. The V-band photometry reported in Table~\ref{tab:data2} for JaFu~1 has been corrected to remove the contamination from the [O~III] emission.

It is possible that JaFu~1 is a member of a small class of PNe having unresolved nebulosities nearly coincidental with the central stars. The prototype of this small class is EGB~6. \citet{Liebert2013} resolved the source of the nebulosity in EGB~6 showing it to be point-like and coincident with a low mass companion at an angular separation of 0.166'', or 96$^{+204}_{-45}$~AU at the estimated distance of 576$^{+1224}_{-271}$~pc. The high electron density of this knot (2.2x10$^6$ cm$^-3$) argues for a disk around the companion to the central star and suggests that the low mass companion has captured some material from the PN ejection process \citep{Bond2016}. If JaFu~1 belongs to this class, then its binary nature would be even more compelling. We note, however, that in the case of EGB~6, the companion is quite distant from the central star, and although an interaction may have taken place, this is unlikely to have been a common envelope pair.

%---------------------
\subsection{The central star of PN JaFu~2}

The mass of the central star of JaFu~2 was calculated by \citet{Jacoby1997} to be  0.55~\msun\  (at a distance of 9.1~kpc) based on a detailed photoionization model. That value of the mass was extrapolated from the stellar evolutionary tracks of \citet{Vassiliadis1994} for their model-based stellar luminosity of 675~\lsun\ and temperature of 100,000 K.  We must scale that luminosity to our currently adopted distance of 11.1~kpc, thereby increasing the luminosity to 1004~\lsun. This change in luminosity, though, increases the CSPN mass by only $\sim$0.005~\msun.

\textnormal{As for JaFu~1, the very low luminosity presents a problem. The \citet{MillerBertolami2016} tracks at Z=0.01 (the metallicity of the cluster is Z=0.0087) suggest that the photoionization model temperature of 100,000~K cannot be reached at the measured low luminosity of 1004~\lsun. The lowest luminosity reached at that temperature is 1860~\lsun for a core mass of 0.53~\msun. Even adopting this mass, the timescale needed to reach these values is 92,000~yrs, again at odds with the low apparent kinematic age ($\sim4900$ yrs) of the nebula (using the expansion velocity from \citet{Pereyra2016}).}

%-----------------

\section{Discussion}

%: I deleted the first parag.
%\textnormal{For this section, we adopt the results based on the evolutionary tracks of \citet{Vassiliadis1994}. While the newer results from \citet{MillerBertolami2016} resolve some issues, others are introduced, and our conclusions are not substantively affected either way.} 

The parameters of these PN systems obtained using the relationships of \citet{Vassiliadis1994} and \textbf{and of \citet{MillerBertolami2016}} are summarized in Table~\ref{tab:data2}. For the four central stars that we know in the Galactic globular cluster system, one of these has a mass that is too high to be explained as the result of normal single star evolution in a globular cluster: K~648 in M15 at 0.57~\msun. This is a good candidate for having arrived at its current stage through a merger channel. In addition, IRAS~18333 may also be too massive to have formed through single star evolution, but the uncertainty in its central star temperature precludes a firm conclusion. 
For JaFu~1 and JaFu~2, the central star masses (0.53 and 0.55~\msun) are consistent with what we expect for normal single star evolution in a globular cluster. Consequently, we do not need to invoke a merger or mass transfer scenario to explain the observed masses. Nevertheless, these low central star masses are predicted to result in slow post-AGB evolution that \textnormal{may} rule out the possibility of observing a PN \textnormal{unless a common envelope interaction has taken place}.

%: I deleted this and summarised it above in half a line.
%This conundrum may be resolved if we invoke a common envelope evolution for these two PNe \citep{Ivanova2013}. In this scenario, a companion interacts with the AGB progenitor of the central star, in-spiraling nearly to the core and ejecting the envelope (which becomes the PN). Because of the fast interaction timescales ($\sim$1 year; \citealt{Passy2012}), the radius of the AGB star is reduced to that of a pre-WD ($\sim$0.1~\rsun) in a short time and the stellar temperature also increases quickly (note that the luminosity is dictated by the core mass). This scenario effectively enables a fast transition time and ionization of the PN gas, even for a low core mass.

%: I have added the following paragraph.
\textnormal{Using the stellar evolutionary tracks of \citet{MillerBertolami2016}, the central star mass of all four stars are in the low range 0.53-0.54~\msun, assuming a low effective temperature for IRAS~18333. However, all the measured kinematic ages of the nebulae are grossly lower than the predicted timescales to reach the measured effective temperature. A common envelope interaction can help to reconcile the timescales, but then the prediction is that all these objects harbor a close binary today. Moreover, we suspect that the low luminosities of the central stars of JaFu~1 and JaFu~2 may indicate a RGB rather than an AGB origin. If so, then their masses could be lower still and a common envelope origin would be even more likely. Finally in the case where \iras\  has a higher effective temperature, its mass of 0.59~\msun\ would be consistent with a past merger and thus, the nebular kinematic age and the post-AGB age could be reconciled.}

For JaFu~1, additional factors have emerged with our {\em HST} observations. The ``necklace''-like nebular morphology \citep{Corradi2011} and the [O~III] and H$\alpha$ emission from the central source offer additional evidence for a post-common envelope evolutionary path.

In the case of JaFu~2, the archive {\em HST} image is mildly reminiscent of that for Ps~1. Except that it exists in a GC, and that its kinematic age is at odds with the predicted post-AGB evolutionary timescales, the arguments for binarity for this PN are the weakest of the group and are not otherwise compelling.

\citet{Buell2012} presented models to illustrate that all of these objects can, in fact, originate with single stars. To do so, they must have been pre-enriched in helium, and therefore must be second or third generation cluster stars. The added helium enables the core to grow in mass by a few crucial hundredths of a solar mass. Those models, though, cannot explain either the high nebular mass in JaFu~1 (0.6 \msun) nor the low nebular mass for \iras\ ($\sim$0.002~\msun) as discussed below. \textnormal{A similar argument applies to the recent evolutionary tracks by \citet{MillerBertolami2016}.}

\subsection{Ps~1, the PN from a merger}

%: I would eliminate this paragraph, we have said this a million times.
%Alves et al. (2000) point out that the nebular carbon abundance of Ps~1 is enhanced, which is usually an indication of enrichment during the third dredge-up. However, only stars more massive than $\sim1$~\msun\  are thought to experience a third dredge-up. A 1~\msun\ progenitor, though, significantly exceeds the turnoff mass for M15, but can be explained naturally as the outcome of a blue straggler or other type of merger \citep{Otsuka2015}.

%: I wold instead write the following premise:
Whether Ps~1 derived from a merger, as we would deduce using the \citet{Vassiliadis1994} evolutionary tracks, or from common envelope evolution, as we would instead conclude from the \citet{MillerBertolami2016} tracks, the nebular morphology deserves some discussion in this context.

%In addition, 
Ps~1 has a morphology that is fairly typical of elliptical PNe with the additional element that it exhibits a prominent arc of emission at one end of its major axis. The arc is likely to be a fast low-ionization emission region or ``FLIER'' \citep{Balick1998}. The {\em HST}  image in [N~II] confirms that the arc is bright in this low ionization line. FLIERs are relatively common in elliptical PNe, although they usually are seen in opposing pairs. A binary central star provides a natural origin for the FLIER phenomenon \citep{Soker1998,Miszalski2009b}. If a companion is present to provide the focus to generate FLIERs, then some remnant of the suspected merger must have survived. 

\subsection{IRAS~18333, a merger, a nova, or something else?}

\iras\  remains a puzzling object. If it truly is a PN, then it has unique characteristics. The principal anomalies are: (i) the nebular composition is definitively H- and He-poor, yet the star certainly contains helium. This anomaly is a glaring incompatibility when associating the nebula to the apparent parent star. Only four other PNe are known to be H-deficient: the central knots within Abell 30 and 78 \citep{Jacoby1983}, the central emission knot within Abell 58 \citep{Seitter1987}, and IRAS~15154-5258 \citep{Manchado1989}. None of these is also known to be He-deficient. (ii) The central star mass (0.56-0.65 \msun) may be substantially higher than expected for a globular cluster post-AGB star under reasonable estimates for its stellar temperature and luminosity.
(iii) The nebula has an extraordinarily high dust-to-gas ratio of $\sim0.3$ \citep{Muthumariappan2013}, where typical values for a PN are 10 to 100 times lower. (iv) The nebular mass (gas plus dust) is very low,  $\sim0.002$~\msun\ \citep{Borkowski1991}, although this mass is consistent with a nebula from which all the hydrogen and helium were removed.

% \citet{Frew2008b} noted how all known post-CE PNe have a surface brightness that is too low for their sizes, indicating that they have low nebular masses, providing another link to a possible binary origin. 

In addition, there are secondary anomalies that are not so extreme as those listed above, but contribute to the overall picture of this object. The number ratio of Ne/O is about 0.8 \citep{Borkowski1991}. This is quite high for PNe where the average ratio is 0.21 \citep{Henry1989} with little dispersion. On the other hand, there is another object, BoBn~1, which is also a ``halo'' PN, that has an even higher Ne/O ratio of 1.63 \citep{Henry2004}. Secondly, the structure of the nebula morphology is highly atypical of PNe, which generally conform to a few representative patterns. In this case, interactions with the ISM have been invoked to explain the unusual appearance \citep{Borkowski1993}, but with the improved  {\em HST}  image, that explanation is less convincing. Lastly, \citet{Monaco2004} reports that the central star exhibits an H$\alpha$ excess, which is a feature seen in very few other PNe (EGB~6, \ja, Tol~26) and may be indicative of a disk.

It is fair to ask whether IRAS~18333 is truly a PN or some other kind of object. There is, though, no other class of object that explains the observed features without introducing a different set of serious problems. Perhaps a neon nova comes closest, where dominant emission lines from neon and oxygen are seen. For a nova, the ejected mass is similar to the \iras\ nebular mass \citep{Gehrz2008}, but the remnant WD would be massive. In that scenario, the source of the neon is from an oxygen-neon WD that descends from a massive ($>6$ \msun) progenitor. High mass progenitors are exceedingly unlikely to exist in an old globular cluster, even in a merger scenario. In fact, the nova option was considered by \citet{Gillett1989} but rejected because the nebular material showed no evidence for fast flows.

While a nova scenario fails to resolve the anomalies, the PN explanation is not much better. If the object were not a member of an old globular cluster where a high mass progenitor can be immediately rejected, the nova option would be  attractive and may ultimately prove to be preferred. Two other hydrogen-deficient PNe, the central emission sources of A30 and A58, have anomalously large abundances of Ne \citep{Wesson2008, Lau2010}. These abundances are at odds with the late thermal pulse scenarios envisaged to explain them \citep{Iben1983,Herwig2001}.

% \citet{Lau2010} sought to explain the observations of V605~Aql with a modified nova model, although only unrealistically finely tuned models could be made to agree with all the observations. 

% Can the nebula be a young disk object in the foreground (galactic latitude is $-7.6$ degrees)? Based on radial velocities, \citet{Gillett1989} argue convincingly that both the nebula and the star are cluster members.

% We can consider another creative explanation; perhaps the star and nebula are unrelated but are spatially coincident for the moment. PN mimics like this (``interlopers'') are well-known to exist \citep{Frew2010} and provide a nice resolution to the first anomaly. However, the stellar and nebular velocities are identical within small uncertainties, and we would still be left trying to explain the source of the H- and He-poor gas. 

\iras\  is so bizarre that we consider its classification as a PN to be highly suspect \citep{Zijlstra2002}. We include it in our discussion, but it should not carry much weight in explaining the presence of PNe in GCs.

As an aside, \citet{Hertz1983} report that the M22 \textnormal{X-ray} source B is about 50 arcsec from \iras. In addition, a rare dwarf nova has been reported in this cluster \citep{Anderson2003}. M22 appears to host a variety of odd phenomena that are associated with interacting binaries.

\subsection{JaFu~1: a post common envelope PN central star }

In this PN the mass of the central star allows for a single star interpretation, while the discrepancy between the kinematic and evolutionary ages indicated by the \citet{MillerBertolami2016} tracks indicates a common envelope origin. There are, however, two additional odd properties to consider. First, the nebula morphology is neither typical nor unique. It is highly structured in a knotty ring-like equatorial formation. The appearance of JaFu~1 is extremely similar to the Necklace PN \citep{Corradi2011} which clearly has a binary central star that has been interpreted as a post-common envelope system. Nebulae in this class are considered to be ejected by common envelope interactions in close binary stars \citep{Miszalski2009}.

Second, \citet{Jacoby1997} derived the ionized mass of the nebular shell to be 0.4 \msun. This value must be revised to $\sim0.6$ \msun\ to account for the more accurate and 22\% larger distance found by \citet{Lee2002}. Combined with the central star mass of 0.55 \msun, the total progenitor mass of $\sim1.1$ \msun\ is incongruously high for a globular cluster star and presents a problem for a single star origin.

We conclude that the progenitor star of JaFu~1 has undergone some mass enhancement. During its evolution, a large fraction of the additional mass was ejected. Consequently, it is a candidate to be classified as a post-common envelope binary in which the companion has played a role in {\bf ejecting} mass, and in the formation and shaping processes. If truly present, a companion also provides an explanation for the excess central [O~III] and H$\alpha$ emission.

% Orsola points out that mass enhancement and CE evolution is difficult to achieve, though not impossible. Perhaps some mass was lost from the companion as well.

\subsection{JaFu~2}

The central star mass of JaFu~2 is small enough to be consistent with the expected evolution for a single star in the host cluster, except if we consider the discrepancy between the kinematic age of the nebula and the presumed evolutionary timescale of the central star according to the \citet{MillerBertolami2016} tracks. Thus, there is no evidence for central star mass augmentation. On the other hand, the morphology of JaFu~2 exhibits subtle similarities to Ps~1. Both are elliptical PNe having an enhanced arc at one end of the major axis. Also, JaFu~2 exhibits a hint of a central torus of emission that is often associated with a binary interaction \citep{Miszalski2009b}. In total, though, the evidence for JaFu~2 having a binary star origin is weaker than for the other GC PNe.

\subsection{Globular cluster PNe among the halo PNe}

These four globular cluster objects are sometimes classified together with about a dozen additional old population objects as ``halo'' PNe. \citet{Howard1997} derived the chemical compositions for nine of these; abundances are available from \citet{Borkowski1991} for \iras\, and from \citet{Jacoby1997} for \ja\  and JaFu~2.

Another halo object is PN~G135.9+55.9 (also known as SBS~1150+599A and TS01). It has been the subject of considerable discussion due to its anomalously low O/H. See \citet{Tovmassian2010} for a summary. The central star of PN~G135.9+55.9 is a close binary having a short 3.924 hr period \citep{Napiwotzki2005}. It is comprised of two hot degenerate stars with masses of $\sim0.86$ and $\sim0.54$ \msun. The masses and period suggest that PN~G135.9+55.9 is a good candidate to become a Type Ia supernova.

BoBn~1 \citep{Boeshaar1977} is also a highly deviant halo PN, and has been associated with the Sagittarius dwarf galaxy \citep{Zijlstra2006}. It shares two unusual properties with the PNe in GCs that are difficult to explain through normal stellar evolutionary paths. First, both BoBn~1 and \iras\  exhibit high Ne/O ratios of $\sim1$, whereas the average ratio for PNe is 0.21 \citep{Henry1989}. \citet{Otsuka2010} proposed a double-$\alpha$ capture mechanism to explain the neon enrichments, a process normally associated with massive progenitor stars. Second, both BoBn~1 and Ps~1 exhibit very high C/O ratios of 13 and 8, respectively, or log (C/H) values of 9.02 and 9.25, respectively.

Halo PN central star masses are not well determined because their distances are generally uncertain. While we can contrast the nebular properties of the halo and GC PNe, we cannot compare their central star masses.

Like the globular cluster system, the halo PNe comprise a small group. One interpretation is that these objects are a highly selected cohort because they derive from a sub-solar progenitor mass and a low metallicity population, a combination that for some reason allows for special circumstances to enable the formation of a PN when one normally cannot be produced. As a group, these PNe have properties that are otherwise rare among the general PN population, again suggesting that they share a similarly rare history (e.g., binarity, common envelope phase, unidentified evolutionary physics).

\section{Conclusions}

The presence of PNe in globular clusters represents a dilemma for stellar evolution unless these unusual objects are the descendants of a binary star phenomenon, such as mass transfer, mergers, or common envelope evolution. We have tested that solution by estimating their remnant central star masses in a more uniform manner than has been considered before, including comparing the results using different stellar evolutionary models. If the progenitor star masses were enhanced, we would have derived masses of the central stars that are higher than WD masses in the globular clusters ($\sim0.53$ \msun).

For Ps~1, the central star mass is higher than the predicted WD mass by $\sim0.05\pm0.02$~\msun. This object is a very good candidate for having a history of mass augmentation through mass transfer or a complete merger.

JaFu~1 and JaFu~2 have masses that are no more than $0.02\pm0.02$~\msun\ higher than the WD expectation value, and so their situation is less clear. For JaFu~1, secondary indicators (morphology, central star emission lines, massive nebula) suggest that a binary interaction is very likely to have been a factor. For JaFu~2, the evidence for binarity is weak - chiefly that it is a PN formed in a GC. The remaining object, \iras\ in M22, has so many peculiarities that its classification as a PN is questionable and binarity may be a plausible way to explain some of its bizarre properties.

% added
\textnormal{If the \citet{MillerBertolami2016} stellar evolutionary tracks are adopted, masses are generally consistent with expectations from old populations, but the kinematic ages of all nebulae are far too short to be consistent with the post-AGB evolutionary timescales of the stars, a discrepancy that could be reconciled if all of these objects are post-common envelope nebulae. In this case a companion should be present, but possibly too faint and of too low mass to be detected.}

Given the small number of objects available for study (three), where about half (1-2) likely require a binary interaction for their formation, a general argument for binarity is far from compelling. Thus, the mechanisms that produce PNe in old populations remain an open question although binary interactions must be a factor in some cases.

This problem would clearly benefit from a larger sample. These objects, though, are so rare in the Galaxy and so difficult to find beyond the Milky Way \citep{Jacoby2013} that resolving this dilemma continues to be a challenge.

% Added for referee's second report
\textbf{While not the primary goal of this paper, we have compared the application of stellar evolutionary models from different authors to evaluate the likelihood of binarity for the PNe in GCs. This has served to demonstrate the broad range of conclusions that one may reach. Despite their long history of use and despite the fact that no overshooting was used, stellar models of \citet{Bloecker1995} and \citet{Vassiliadis1994}, for example, are still widely adopted. Often in the PN field, an evolutionary model is used to interpret observations, but those interpretations are not contrasted with results using other models. Yet, important characteristics (e.g., core mass, luminosity, transition time) vary significantly depending on the choice of overshoot prescription, criteria to terminate the AGB phase, and mass-loss rate during the transition time that go into the models.  Those choices are based on subtle reasoning that is lost when the models are applied. Our comparisons here highlight a few aspects of the diversity of conclusions that one can reach when using different models.}

\acknowledgments

We wish to thank the anonymous referee for suggestions that significantly improved our interpretations and the presentations in this paper, \textbf{and for the insights summarized in the final paragraph}. Partial support for program GO-11558 was provided by NASA through a grant from the Space Telescope Science Institute, which is operated by the Association of Universities for Research in Astronomy, Inc., under NASA contract NAS 5-26555.

During the course of this work, support was provided by the following institutions: WIYN Observatory \textnormal{and Giant Magellan Telescope Organization} (Jacoby), American Museum of Natural History (De Marco), and California Institute of Technology (Davies).

L. Lahm assisted with the initial {\em HST} photometry as a student intern at the American Museum of Natural History.


%% in parentheses, after the keyword, but they will not be verified.

%% {\it Facilities:} \facility{Nickel}, \facility{HST (STIS)}, \facility{CXO (ASIS)}.

\software{IRAF}

\facility{{\em HST}}

\clearpage

\begin{table}
\caption{{\em HST} imaging data for IRAS~18333 and JaFu~1}

\begin{tabular}{llccrr}
\hline
PN Name     &    Date         &  Camera  &  Filter & Exposure, s &  Proposal ID\\
\hline
IRAS~18333  &  1994 Apr 07  &  WFPC2   &  F502N  & 3600        &   5404 (Harrington) \\ 
IRAS~18333  &  2010 Mar 02  &  ACS     &  F502N  & 5306        &  11558  (De Marco)\\
JaFu~1      &  2010 Mar 14  &  ACS     &  F502N  & 2388        &  11558  (De Marco)\\
JaFu~1      &  2008 Mar 14  &  WFPC2  &  F555W  &  320        &  11558  (De Marco)\\
JaFu~1      &  2008 Mar 14  &  WFPC2  &  F841W  &  320        &  11558  (De Marco)\\
JaFu~1      &  2008 Mar 14  &  WFPC2  &  F656N  & 1000        &  11558  (De Marco)\\
\hline
\end{tabular}
\label{tab:obs}
\end{table}

\begin{table}
\caption{Characteristics of the four host GC}

\begin{tabular}{llrrrr}
\hline
PN Name       & Host     & Distance$^a$   & Age$^b$           & Metallicity$^c$   & Radial Velocity$^d$\\
              & GC       &  (kpc) &(Gyr)      & [Fe/H]    & \kms  \\
\hline                
Ps~1 (K~648)  & M15      & 10.0       & $13.2\pm1.0$  & $-2.4$        & $-107.0$ \\
IRAS~18333    & M22      & 3.1        & $13.1\pm1.2$  & $-1.7$        & $-146.3$ \\
JaFu~1        & Pal~6    & 7.2        & $\sim11.7$    & $-1.2$        & $+180.6$ \\
JaFu~2        & NGC~6441 & 11.1       & $11.6\pm1.3$  & $-0.6$        & $+16.5$ \\
\hline
\end{tabular}
\tablecomments{$^a$M15 from \citet{McNamara2004}, M22 from \citet{Monaco2004}, Pal~6 from \citet{Lee2002}, NGC~6441 from \citet{Pritzl2001}; $^b$ M15 from \citet{McNamara2004}; M22 and NGC~6441 from \citet{Marin-Franch2009} relative to M15; Pal~6 from \citet{Lee2002}$^c$ M15 from \citet{Dotter2010}; Pal~6 from \citet{Lee2002}; M22 and NGC~6441 from \citet{Marin-Franch2009}; $^d$ M15, M22, and NGC~6441 from \citet{Harris1996} (2010 edition), and Pal~6 from \citet{Lee2004}}
\label{tab:data1}
\end{table}

% REVISE TABLE TO SHOW Teff, Lsun, and Mass
\begin{table}
\caption{Characteristics of the four GC PN}

\begin{tabular}{lllllcll}
\hline
PN Name          & CS $V$  & CS $I$ & \textbf{CS Mass (VW)} & \textbf{CS Mass (MB)}   & Progenitor       & PN diam.  &    PN \\
                 & (mag)  & (mag) & (\msun)        & (\msun) & Mass$^a$ (\msun) & (pc) &  type \\
\hline                
Ps~1 (K~648)$^b$ & 14.73  & 14.93 & $0.57\pm0.01$  & \bf{$0.53$} & 1.2              &  0.14       &   elliptical; H-normal\\
IRAS~18333$^c$   & 14.53  & 14.25 & $0.56-0.65$    & \bf{$0.54-0.59$} & 1.0              &  0.11       &   irregular; H-deficient\\
JaFu~1$^d$       & 22.40  & 19.70 & $0.53\pm0.02$  & \bf{$0.53$} & 0.8              &  0.26       &   ring, knots; H-normal \\
JaFu~2$^e$       & 19.91  & 19.72 & $0.55\pm0.02$  & \bf{$0.53$}  & 1.0              &  0.25       &   elliptical; H-normal \\
\hline
\end{tabular}
\tablecomments{$^a$From the IFMR relation of Weidemann (2000), assuming normal single star evolution produced the CS mass. $^b$Alves et al. 2000. $^c$Monaco et al. 2004. $^d$This paper; V mag has been corrected for contamination from the central star's [O~III] emission. $^e$Jacoby et al. 1997. All magnitudes are on the Vega system. \textbf{ ``VW'' CS masses are derived using the \citet{Vassiliadis1994} tracks while ``MB'' CS masses are based on \citet{MillerBertolami2016} tracks.}}
\label{tab:data2}
\end{table}

\clearpage

\bibliographystyle{apj}                       %% AASTeX
%\bibliography{biblio,../../BibliographyFiles/bibliography}
\bibliography{JaFu1Paper_v8}

\end{document}